# An EEG study of creativity in expert classical musicians

Tom De Smedt[1], Lieven Menschaert, Pieter Heremans, Ludivine Lechat, Gaëlle Dhooghe

ABSTRACT – Previous research has shown positive correlations between EEG alpha activity and performing creative tasks. In this study, expert classical musicians (*n*=4) were asked to play their instrument while being monitored with a wireless EEG headset. Data was collected during two rehearsal types: (a) in their regular, fixed ensemble; (b) in an improvised, mixed ensemble with unfamiliar musicians and less rehearsal time. A positive correlation was found between alpha power and the improvised setup ($p<0.01$, $d=0.4$). A positive correlation was also found between alpha power and more intense play ($p<0.01$, $d=0.2$). There was a negative correlation between alpha power and arousal due to stress, e.g., frowning after playing a false note ($p<0.01$, $d=0.6$). Finally, the real-time capabilities of wireless EEG monitoring were explored with a data visualisation during live performance on stage.

## 1    Introduction

Creativity is a multi-faceted phenomenon (Veale, Feyaerts, and Forceville 2013) with, as of yet, no single best definition. Is it a mental process, an ability, an activity, a product, or all of these? The study of creativity can be traced to J. P. Guilford's divergent thinking tests (Guilford 1956), where subjects are asked to come up with as many solutions as possible to an open-ended problem. For example: 'what can you do with a paperclip?' Variations of divergent thinking tests are still widely used to assess creativity (e.g., Torrance 1988) but it has also been argued that really they assess the ability to come up with *potentially* creative ideas (Runco 1991), which is not the same.

Since Guilford, more than fifty years of psychological research has shown that creativity involves intelligence (Guilford 1967), intrinsic motivation (Amabile, Hill, Hennessey, and Tighe 1994), expertise (Ericsson 1998), knowledge (Weisberg 1999), gatekeeping (Csikszentmihalyi 1999), and personality traits such as openness to experience (Martindale 1989) and introversion (Feist 1998). For an overview, see Sternberg (1999).

A recent area of research is computational creativity. It explores computational approaches to simulate and evaluate creativity, using AI techniques such as semantic networks (Boden 2004), frames (Fauconnier and Turner 2008) and machine learning (Mordvintsev, Olah, and Tyka 2015). Experiments in this field are diverse and interdisciplinary – since there is no single best definition of creativity from which to start. For an overview, see De Smedt (2013).

The study of creativity has also attracted interest from neuroscience, using methods such as EEG and fMRI. EEG or electroencephalography records the brain's electrical activity resulting from nerve impulses, by placing electrodes along the scalp. Usually, the EEG signal is filtered into four frequency bands for analysis: a delta wave (0.5-4 Hz) and a theta wave (4-8 Hz) associated with sleep, an alpha wave (8-13 Hz) associated with relaxation, and a beta wave (13-30 Hz) associated with alertness.

---

[1] tom@organisms.be

For a comprehensive overview, see Teplan (2002). Several studies report a correlation between alpha waves and creative tasks (Fink and Benedek 2014). For example, when solving a divergent thinking test, the subjects with a higher test score also show higher alpha power (Martindale and Mines 1975; Jaušovec 2000; Bazanova and Aftanas 2008). In another study with word association tests, subjects show higher alpha power during the 'Aha!' moment, i.e., when coming up with an insightful solution (Jung-Beeman et al. 2004). A study with dancers shows that professional dancers who imagine themselves performing an improvised dance show higher alpha power than novices (Fink, Graif, and Neubauer 2009).

In our study, we would like to know if musicians are more creative when improvising. Intuitively, many musicians describe a 'flow' during which they are pleasantly immersed in their play. Flow is defined by Csikszentmihalyi (1991) as a mental state of complete absorption (e.g., reduced self-consciousness and sense of time) and peak creative performance; the 'sweet spot' between a task that is too easy and dull, and a task that is too difficult and frustrating. Flow has been observed in improvising soloists (Parncutt and McPherson 2002), and has been linked to lower heart rate in piano players (de Manzano, Theorell, Harmat, and Ullén 2010), and to higher alpha power (De Kock 2014).

## 2 Experimental Setup

The study was conducted in collaboration with the Festival of Flanders for classical music. Each year, they organize a 10-day 'pressure cooking' event[2]. Each day, musicians are drawn from different ensembles and orchestras, and then mixed to rehearse a music piece once or twice. The music piece is then performed live on stage later that day. This improvisational approach (hereafter MIXED) differs from a regular rehearsal, which is generally repeated multiple times in a fixed ensemble (hereafter FIXED) with, for example, more time to find the right balance or rehearse complicated parts. In both scenarios, the musicians will want to deliver a good live performance.

The goal of the FIXED and MIXED rehearsal is the same (a high-quality live performance), but the MIXED rehearsal is arguably more challenging creatively, with less rehearsal time and unfamiliar co-musicians. With this in mind, we hypothesize that musicians will play at peak performance in the MIXED rehearsal, and thus show higher EEG alpha power.

### 2.1 Participants

Four musicians, respectively playing the OBOE, VIOLIN, BASS, and PIANO, were interested to participate in the experiment. All of them are professionals, and known classical musicians, aged 40-50. The VIOLIN player is female. It is useful to include musicians that play different instruments, to observe if results are also consistent across different instruments. To illustrate this, the oboe is popularly considered to be a 'difficult' instrument (*embouchure* and air flow), the violin bass is heavier and requires its player to stand, making more pronounced body movements that could introduce noise in the EEG signal, and so on.

---

[2] http://www.b-classic.be

2.2     Equipment

The EEG data was captured using a wireless headset (Patki, Grundlehner, Nakada, and Penders 2011) developed by Imec[3] for research purposes. As shown in Figure 1, the headset has four dry electrodes, located at C3, C4, Cz and Pz according to the 10-20 international system of electrode placement (Klem, Lüders, Jasper, and Elger 1999). The electrodes transmit a raw signal to the computer by Bluetooth (i.e., short-wavelength radio). Two musicians wearing the device are shown in Figure 2.

We monitored the signal from electrode Cz (the top middle of the head) at 256 samples per second, using a Fast Fourier Transform (FFT) to derive the EEG power spectrum (alpha, beta, etc.) The spectral power values are relative (i.e., delta, theta, alpha and beta values sum to 1.0). Our algorithm in Python code is shown in Figure 3. It is based on the open source PyEEG toolkit (Bao, Liu, and Zhang 2011), with the addition of a bandpass filter and a 10-second Hanning window to smoothen the band waves.

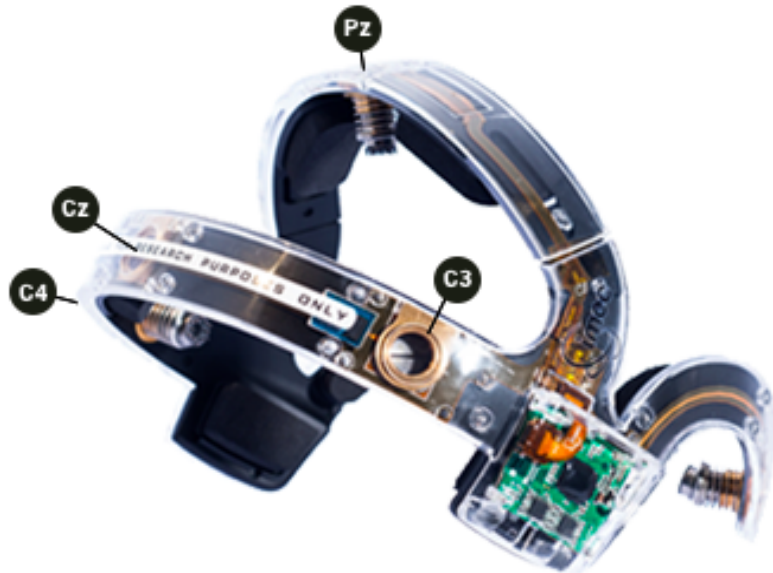

Figure 1. Wireless EEG headset developed by Imec.
© Imec/Holst Centre. Used with permission.

---

[3] http://www.imec.be

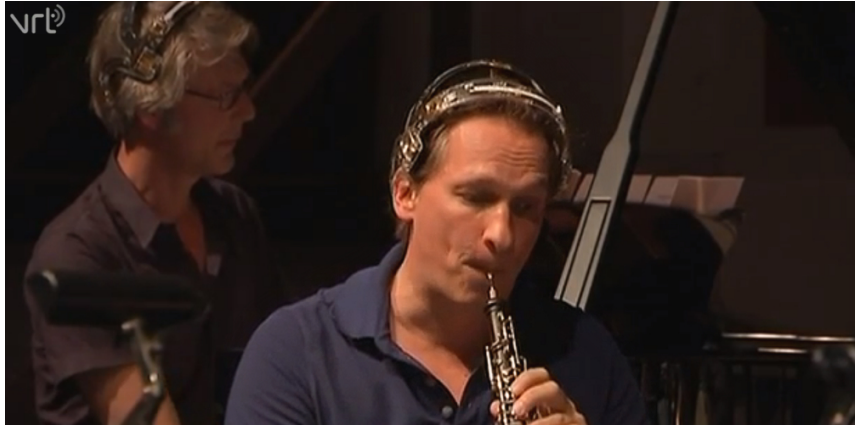

Figure 2. Wireless EEG headset worn by the PIANO and OBOE players.
© VRT. Used with permission.

```python
from numpy import hanning, fft, floor

from scipy.signal import butter, lfilter

def bandpass(signal=[], lo=0.5, hi=30, fs=256, order=4):
    f = butter(order, [lo * 2.0 / fs, hi * 2.0 / fs], 'band')
    f = lfilter(f[0], f[1], signal)
    return f

def power(signal=[], hz=[0.5, 4, 8, 13, 30], fs=256, relative=True):
    p = []
    a = bandpass(signal)
    a = a * hanning(len(a))
    a = fft.fft(a)
    a = abs(a)
    n = len(a) / float(fs)
    for i in xrange(len(hz)-1):
        x = floor(hz[i+0] * n)
        y = floor(hz[i+1] * n)
        s = a[x:y]
        p.append(sum(s))
    if relative:
        p = [v / sum(p) or 1 for v in p] # sum to 1
    return p
```

Figure 3. EEG spectral analysis in Python programming code.

## 2.3 Data Collection

The experiment then proceeds as follows. We visited each musician during a FIXED rehearsal with their regular ensemble, at their home or at their academic institute. For each musician, we recorded approximately 15 minutes of EEG alpha power, using the hardware and software outlined above. Recording starts halfway during the rehearsal, to exclude data of the musician warming up or winding down. A few weeks later, we revisited each musician during a MIXED rehearsal with unknown musicians, at the concert venue of the Festival of Flanders event. For each musician, we again recorded approximately 15 minutes of EEG alpha power. During the MIXED rehearsals, the weather was unusually hot in the region[4].

## 2.4    Data Annotation

Each session was also recorded on video, and each video timeline was annotated in 10-second intervals with events such as 'playing' or 'moving', for cross-reference. At a glance, the FIXED rehearsals progress in a more relaxed fashion, with more discussion among musicians, more retakes, and also more laughs. On the other hand, after the MIXED rehearsals two musicians intuitively reported an 'Aha!' moment, i.e., a moment during rehearsal when they discovered an interesting balance among musicians, or the prospect of an excellent live performance.

# 3    Evaluation

Table 1 shows the average alpha power and standard deviation for each musician during each FIXED rehearsal (i.e., regular ensemble) and each MIXED rehearsal (i.e., improvised ensemble). Figure 4 shows alpha power relative to the entire spectrum. For three musicians, alpha power increases during MIXED rehearsals (+5-30%). For one musician, alpha power decreases (−9%).

The difference in alpha power between FIXED and MIXED rehearsals was tested with a $4 \times 2$ ANOVA[5] (4 musicians, 2 rehearsals, $15 \times 60 \times 256$ samples per musician per rehearsal) and is statistically significant ($p<0.001$). The effect is fairly small (Cohen's $d=0.4$, where $d \geq 0.2$ is small, $d \geq 0.5$ medium, $d \geq 0.8$ large). On average, alpha power increases by 15% during MIXED rehearsals.

A goodness of fit test[6] shows that alpha power is not normally distributed for any musician. Since ANOVA (analysis of variance) assumes normality, each sample was first normalized with `log(x/(1-x))` as recommended in Gasser, Bächer, and Möcks (1982).

|        | FIXED REHEARSAL | | MIXED REHEARSAL | | DIFFERENCE |
|--------|------|------|------|------|------|
|        | $M_1$ | $SD_1$ | $M_2$ | $SD_2$ | $M_2 / M_1 - 1$ |
| OBOE   | 0.095 | 0.025 | 0.100 | 0.030 | +5% |
| VIOLIN | 0.125 | 0.030 | 0.115 | 0.025 | −9% |
| BASS   | 0.100 | 0.025 | 0.130 | 0.035 | +30% |
| PIANO  | 0.150 | 0.060 | 0.195 | 0.015 | +30% |

Table 1. Mean difference in alpha power between rehearsals in
regular (fixed) ensembles and in improvised (mixed) ensembles.

---

[4] http://www.weather.com/forecast/news/europe-heat-wave-record-highs-june-july-2015
[5] http://vassarstats.net/anova2u.html
[6] http://www.clips.ua.ac.be/pages/pattern-metrics#ks2

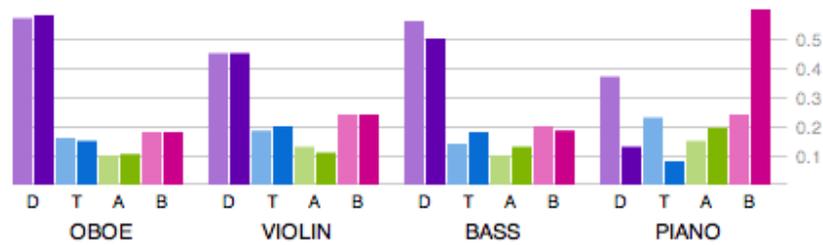

Figure 4. EEG power spectrum for each musician during rehearsal in regular, fixed ensemble (light) and in improvised, mixed ensemble (dark).

Note the unusual increase in beta power for the PIANO player. One explanation is the intense hand-eye coordination involved, i.e., play piano, watch partiture, quickly turn page, quickly continue playing.

3.1   Evaluation of Personality

Why does alpha power decrease for the VIOLIN player? One explanation is gender. Gender-related EEG differences during creative tasks have also been reported in Razumnikova (2004). Another explanation is signal noise related to gender physiology. The other three musicians (males) have less hair that can interfere with the wireless headset's dry electrodes, which need to be resting firmly on the scalp. Finally, a third explanation is personality. Given that some individuals are more extraverted while some are more introverted (Eysenck 1967), introverted musicians may consider the improvised setup to be more demanding (i.e., more cortical arousal and less alpha), as social interaction detracts them from reaching peak performance. However, the relation between personality and EEG is a topic of discussion, see for example Matthews and Amelang (1993), and Fink and Neubauer (2008).

Each of the participating musicians was asked to complete a self-reported, online[7] Big Five personality traits questionnaire (Goldberg 1990). The results for the Extraversion and Openness to Experience percentile scores are summarised in Table 2. Table 2 shows that the VIOLIN player is more introverted.

|        | AGE   | GENDER | E      | O      |
|--------|-------|--------|--------|--------|
| OBOE   | 40-50 | m      | ≥ 75%  | ± 50%  |
| VIOLIN | 40-50 | f      | ≤ 25%  | ≥ 50%  |
| BASS   | 40-50 | m      | ≥ 75%  | ≤ 50%  |
| PIANO  | 40-50 | m      | ≥ 75%  | ≥ 50%  |

Table 2. Big Five percentile scores for Extraversion (E) and Openness to Experience (O).

---

[7] http://personality-testing.info/tests/BIG5.php

## 3.2 Evaluation of Timeline Events

Each FIXED and MIXED session was recorded on low-quality video. Two reviewers independently annotated each video timeline at 10-second intervals with relevant events, for cross-reference:

1. PLAYING    musician appears to be playing,
2. MOVING     musician appears to be moving (e.g., nodding, talking),
3. AROUSED    musician appears to be nervous (e.g., frown, grimace, stop),
4. PLEASANT   reviewer likes the music being played.

With multiple reviewers, we can calculate inter-rater agreement using Fleiss' kappa. The kappa score represents a statistical degree of consensus as a value between −1.0 and +1.0, where $k < 0$ indicates poor consensus among the reviewers. Not surprisingly, there is agreement whether musicians are playing ($k$=+0.5) or moving ($k$=+0.2). But there is little or no agreement whether musicians appear to be aroused ($k$=0.0), or whether the performance is enjoyable ($k$=−1.0).

Additionally, one reviewer annotated the parts where a musician's play was more INTENSE (e.g., faster tempo, solo).

A snapshot of annotated PIANO data is shown in Figure 5, illustrating how EEG data can then be examined in relation to events (e.g., is alpha power related to arousal?). Sections 3.2.1-6 report the results of six ANOVA-tests on events.

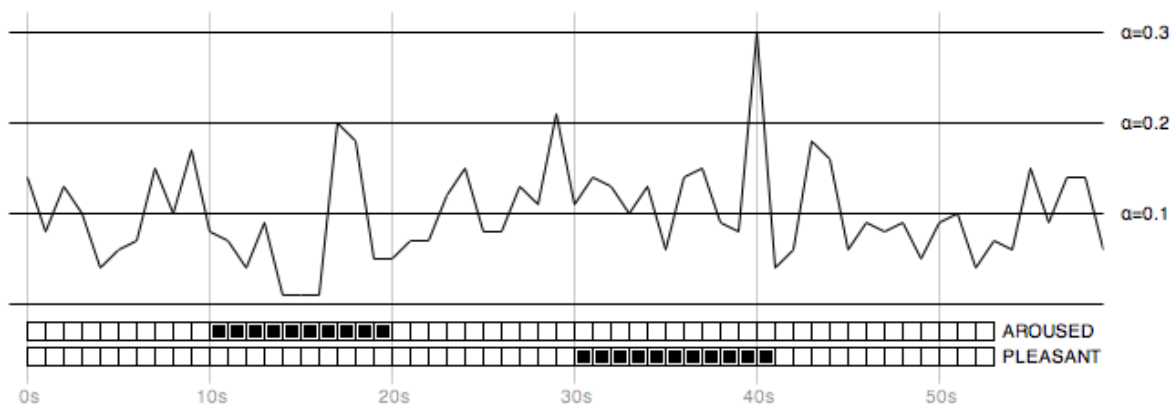

Figure 5. EEG alpha power on an annotated timeline.

### 3.2.1 Alpha and Playing Music

The difference in alpha power when musicians are PLAYING or not is significant in both FIXED ($p$<0.01, $d$=0.2) and MIXED rehearsals ($p$<0.01, $d$=0.1). The effect is very small. On average, alpha power increases by 7.5% when playing.

### 3.2.2 Alpha and Noise

The difference in alpha power when musicians are MOVING (e.g., nodding in tune) or not is significant in both FIXED ($p<0.01$, $d=0.1$) and MIXED rehearsals ($p=0.03$, $d=0.1$), but the effect is trivial. Imec's headset is a research prototype designed for ambulatory monitoring of patients. The design focuses on long battery life, and easy setup by using wireless technology and dry electrodes. The main drawback of dry electrodes is that when the wearer nods his or her head, signal noise can be introduced. The test indeed shows that there is signal noise, but that it is no cause for concern ($d < 0.2$).

### 3.2.3 Alpha and Arousal

The difference in alpha power when musicians are AROUSED (e.g., frowning) or not is significant in both FIXED ($p=0.03$, $d=0.4$) and MIXED rehearsals ($p<0.01$, $d=0.6$). The effect is medium. On average, alpha power decreases by 35% when aroused. One explanation is that playing difficult parts in the partiture requires more attention and more conscious effort, with less room for creative improvisation. The musicians also report that when they play a false note, they are instantly out of the flow and revert to their years of experience.

### 3.2.4 Alpha and Valence

The difference in alpha power during parts of the rehearsal that one or both annotators found PLEASANT and other parts is not significant ($p>0.99$). In any case, the opinion of the annotators as to what is 'good' music is subjective.

### 3.2.5 Alpha and Intensity

The difference in alpha power during parts of the rehearsal that are more INTENSE and other parts is significant in both FIXED ($p<0.01$, $d=0.2$) and MIXED rehearsals ($p<0.01$, $d=0.2$). The effect is small. On average, alpha power increases by 12% during more intense parts.

### 3.2.6 Alpha and Brain Hemisphere

The difference in alpha power between left brain hemisphere (i.e., electrode C3) and right brain hemisphere (C4) is significant in both FIXED ($p=0.02$, $d=0.1$) and MIXED rehearsals ($p<0.01$, $d=0.6$). On average, alpha power is 10% higher in the right brain hemisphere. The effect is larger in MIXED rehearsals (+18%) than in FIXED rehearsals (+4%).

### 3.3 Summary

We compared EEG alpha power (which has been linked to creativity) of expert musicians rehearsing in their comfort zone to rehearsals in an improvised setup. Alpha power correlates positively with the improvised setup. Alpha power also correlates positively with more intensive play, and negatively with arousal due to stress. Higher alpha power is observed in the right brain hemisphere.

## 4      Real-time Visualisation

At the end of each day during the Festival of Flanders event, the musicians performed their rehearsed MIXED piece in front of a live audience (± 500). We wanted to know how well our EEG setup responds in real-time. A projection screen was placed on stage behind the performing ensemble, displaying a real-time visualisation that responds to the alpha power of a musician wearing a headset. This approach is also known as biofeedback (Bersak et al. 2001; Gilleade, Dix, and Allanson 2005), where a video game or multimedia installation is controlled by behavioral or physiological input such as gestures (Jordà et al. 2007) or brain waves (Verle 2007) instead of traditional game controllers. For an overview, see Gurkok and Nijholt (2013).

Similar to previous work (De Smedt and Menschaert 2012), the visualisation consists of an agent-based system (cf. Reynolds 1999) where individual agents (i.e., on-screen elements) interact according to EEG signals. The aesthetic look & feel is based on a neural network (the morphological structure of the brain) and symmetrical patterns emerging from vibrations (Waller and Chladni 1961). Agents will grow in size, connect, and become more symmetrical in shape when alpha power is higher. This results in a rich variety of compositions. A number of possible variations of individual agents is shown in Figure 6. A climatic moment of high alpha power for the BASS player is shown in Figure 7.

We did not encounter any latency problems. The wireless Bluetooth connection did fail on one occasion. This could be attributed to the distance (±15m) between the stage and the PA booth where the receiving computer was located. When this occurred, we reverted to a simulated, randomly generated EEG signal so as not to detract the audience from the live experience.

The technical setup is as follows. One musician on stage is wearing a wireless EEG headset. The headset communicates with a computer that derives the EEG power spectrum from the raw electrode signal (Cz). The alpha power value is sent to a web browser using the bidirectional WebSocket protocol (Fette and Melnikov 2011). The web browser generates the visualisation.

In-browser data visualisation is interesting to explore applications for remote monitoring, e.g., where the physician receives real-time updates on a mobile device of ambulant patients wearing an EEG headset. In-browser graphics rendering based on HTML5 canvas (Fulton and Fulton 2013) works on multiple devices. We first experimented with p5.js[8] before deciding on Three.js (Cabello 2010), an open source JavaScript framework for 2D and 3D graphics that leverages HTML5 canvas and WebGL technologies. The Three.js API has post-processing capabilities useful for fine-tuning the aesthetic look & feel. In this case, a two-pass Gaussian blur filter, a vignette shader[9], and interactive controls for zooming and panning were added.

---

[8] http://p5js.org
[9] http://alteredqualia.com

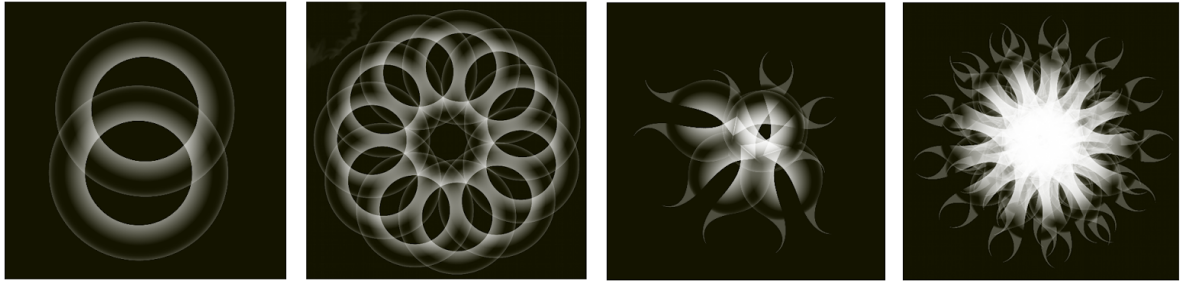

Figure 6. Variations of individual agents, where the complexity is based on alpha power.

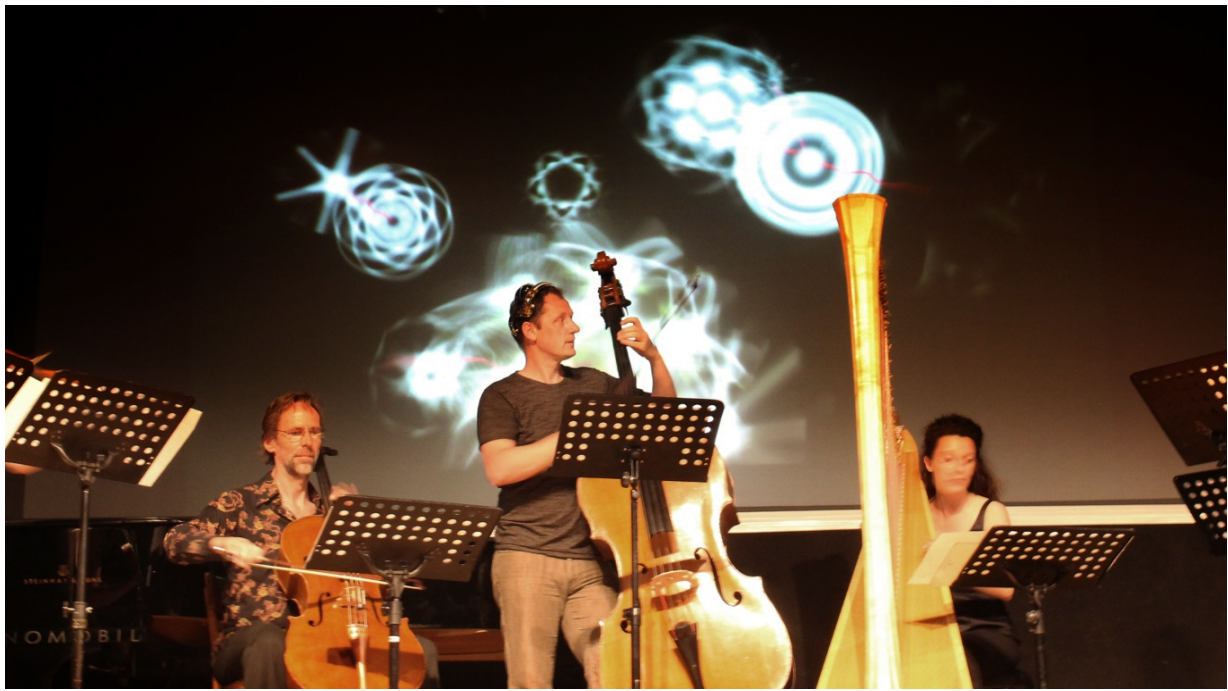

Figure 7. Agent-based visualisation reaches climax as alpha power of the BASS player rises.

A missed opportunity is that we did not offer the audience a chance to vote on climatic moments (e.g., by mobile device or by voting button). Such crowd-sourced data could have been used for analysis.

For entertainment, we did offer the audience a chance to wear an EEG headset during the concert. In this case, the signal was sent to a home automation server[10] using the UDP protocol. The home automation server then controlled the ambient light of ten 3D-printed design lamps[11] scattered across the audience.

---

[10] http://www.loxone.com
[11] https://mgxbymaterialise.com

# 5 Future Work

It is useful to repeat the experiment with more classical musicians (e.g., 50) to replicate the results, for example involving a university orchestra. Although university orchestras are not composed of professional musicians, most members have longstanding expertise. Usually these are undergraduate and graduate students open to the idea of scientific study, and more easily available than professional classical musicians.

# 6 Acknowledgements

The research was funded by the Creative Industries programme (CICI) of the Institute for Innovation by Science and Technology in Flanders (IWT), and conducted in collaboration with the Festival of Flanders for classical music and Imec.